\documentclass[11pt]{article}
\usepackage{moriond,epsfig}

\def\Journal#1#2#3#4{{#1} {\bf #2}, #3 (#4)}

\def\NPB{{\em Nucl. Phys.} B}

\def\PRL{\em Phys. Rev. Lett.}
\def\PRD{{\em Phys. Rev.} D}

\def\JPG{{\em J.Phys.} G}
\def\JINST{{\em JINST}}
\def\PoS{{\em PoS}}

\def\be{\begin{equation}}
\def\ee{\end{equation}}
\def\bea{\begin{eqnarray}}
\def\eea{\end{eqnarray}}

\newcommand{\Dsm}{$D^-_s$}
\newcommand{\Dm}{$D^-$}

\newcommand{\BsDp} {\ensuremath{B^0_s \to D^-_s \pi^+}}
\newcommand{\BdDp} {\ensuremath{B^0 \to D^- \pi^+}}
\newcommand{\BsDK} {\ensuremath{B^0_s \to D^\mp_s K^\pm}}
\newcommand{\fsfdt}{\ensuremath{f_s/f_d}}

\begin{document}
\vspace*{4cm}
\title{Studies of the decay $B^{0}_{s} \to D^{\mp}_sK^{\pm}$}

\author{B. Storaci$^*$ on behalf of the LHCb collaboration }
\address{NIKHEF, \\
        Science Park 105, 1098 XG Amsterdam, The Netherlands\\
        $^*$E-mail: Barbara.Storaci@cern.ch
}

\maketitle

\abstracts{The decay mode $B^{0}_{s} \to D^{\mp}_sK^{\pm}$ 
allows for one of the theoretically 
cleanest time dependent measurements of the CKM angle $\gamma$. 
This contribution reports the world best branching fraction of this decay 
relative to the Cabibbo--favoured mode $B^{0}_{s} \to D_{s}^{-}\pi^{+}$
based on data sample of 0.37 $\rm fb^{-1}$ proton--proton collisions at 
$\sqrt{s} = 7$~TeV collected with the LHCb detector in 2011, resulting in
$BR(\BsDK) = (1.90 \pm 0.12^{stat} \pm {{0.13^{syst}}^{+0.12}_{-0.14}}^{f_{s}/f_{d}})\times 10^{-4}$.
}

\section{Motivation}\label{s:motivation}
The least precise direct measured parameter of the
unitary triangle is the angle $\gamma$. The high abundance of $b \overline{b}$ 
pairs, together with an excellent proper time resolution, 
an excellent particle identification and trigger capability to 
select hadronic final states, 
allows the LHCb experiment to determine this parameter through
a time dependent analysis using the $B^{0}_{s} \to D^{\mp}_sK^{\pm}$ decay. 
Unlike the flavour-specific decay $B^{0}_{s} \to D_{s}^{-}\pi^{+}$, 
the Cabibbo-suppressed decay $B^{0}_{s} \to D^{\mp}_sK^{\pm}$ proceeds 
through two different tree-level amplitudes of similar strength.

These two decay amplitudes can have a large $C\!P$-violating 
interference via $B^0_s-\bar{B}^0_s$ mixing, allowing the determination of 
the CKM angle $\gamma$ 
with small theoretical uncertainties through the measurement of tagged 
and untagged time-dependent 
decay rates to both the $D^-_s K^+$ and $D^+_s K^-$ final 
states \cite{Fleischer}. 
Although the $B^{0}_{s} \to D^{\mp}_sK^{\pm}$ decay mode has been observed 
by the CDF~\cite{Aaltonen:2008rw} 
and BELLE~\cite{BelleBsDspi} collaborations, at present its branching fraction 
is known with an uncertainty around 23\% \cite{PDG}.
Moreover, only the LHCb experiment has both the necessary decay time resolution 
and access to large enough signal yields to perform the time-dependent $C\!P$ measurement.

\section{The LHCb experiment}\label{s:lhcb}
The LHCb detector~\cite{DetectPaper} is a single-arm forward
spectrometer covering the pseudo-rapidity range $2<\eta <5$, designed
for studying particles containing $b$ or $c$ quarks. 
The detector includes a high-precision tracking system (silicon and straw tube technologies) and a dipole
magnet with a bending power of about $4{\rm\,Tm}$. 
The tracking system has a momentum resolution
$\Delta p/p$ that varies from 0.4\% at 5 GeV 
to 0.6\% at 100 GeV,
an impact parameter resolution of 20$\mathrm{\mu m}$ for tracks with high
transverse momentum, and a decay time resolution of 50~fs. 
Charged hadrons are identified using two 
ring-imaging Cherenkov detectors. Calorimeter and muon systems provide 
the identification of photon, electron, hadron and muon candidates.

The analysis is based on a sample of $pp$ collisions corresponding 
to an integrated luminosity of 0.37~fb$^{-1}$,  
collected at the LHC in 2011 at a centre-of-mass energy 
\mbox{$\sqrt{s} = 7$ TeV}.

\section{Selection} \label{s:selection}
The channels considered as signal in this document are the decays \BdDp, \BsDp~ 
and the \BsDK. These decays are all characterized by a similar topology and therefore the
same trigger, stripping and offline selection are used to select them, minimizing
the efficiency corrections. 

The LHCb trigger consists of a hardware stage, based
on information from the calorimeter and muon systems, followed by a
software stage which applies full event reconstruction.

The decays of $B$ mesons can be distinguished from the
background by using variables such as the $p_T$ and impact
parameter $\chi^2$ of the $B$, $D$, and the final state particles with
respect to the primary interaction. In addition, the vertex
quality of the $B$ and $D$ candidates, the $B$ lifetime, and the
angle between the $B$ momentum vector and the vector
joining the $B$ production and decay vertices are used in
the selection. In order to remove charmless background 
a requirement in the flight distance $\chi^2$ of the $D^-_s$ from the $B^0_s$ 
is applied \cite{DsKpreprint}.

Further suppression of combinatorial backgrounds is achieved
using a gradient boosted decision tree
technique~\cite{TMVA}.  
The optimal working point is evaluated directly from a sub-sample of \BsDp~events in data
selected using particle identification and trigger requirements.
The chosen figure of merit is the significance of the \BsDK~signal, scaled according
to the Cabibbo suppression relative to the \BsDp~signal, with respect to the combinatorial background.
Multiple candidates occur in about $2\%$ of the events and in such cases
a single candidate is selected at random.

Particle identification (PID) criteria serve two purposes: separate the Cabibbo-favoured from the 
Cabibbo-suppressed modes (when applied to the bachelor particle) and suppress the misidentified 
backgrounds which have the same bachelor particle (when applied to the decay products of the \Dsm~or \Dm).
All PID criteria are based on the differences in log-likelihood (DLL) between
the kaon, proton, or pion hypotheses. Their efficiencies are obtained
from calibration samples of $D^{*+} \to (D^0 \to K^- \pi^+) \pi^+$ and 
$\Lambda\to p\pi^-$ signals, which are themselves
selected without any PID requirements. These samples are 
split according to the magnet polarity, binned in momentum and $p_{\rm T}$,
and then reweighted to have the same momentum and $p_{\rm T}$
distributions as the signal decays under study.

\section{Mass fit} \label{s:fit}
The three signal decays are distinguished with 
particle identification requirements applied at the final stage of the 
analysis. The signal yields are obtained from extended maximum likelihood unbinned fits to the
data. In order to achieve the highest sensitivity, the sample
is fitted separately for the magnet up and down data. 
The signal line shapes are taken from simulated signal events. 
A mass constraint on the $D_{(s)}$ meson mass is used in order to improve the $B$ mass resolution.

The shape of the signal mass distribution is obtained fitting a double Crystal Ball function 
which consist of a common Gaussian with two exponential tails, one to describe the radiative 
tail present in the lower mass region and the other one describing the higher mass region where only 
the detector resolution is involved.

A common signal shape describes properly both polarities so a simultaneous fit with a
common mean and width of the double crystal ball function is used.
The mean is free to float in all the fits, while the width is fixed in the $B^0_s$-modes 
from the result obtained in data in the \BdDp~fit corrected for the $B^0-B^0_s$ 
differences observed in the simulation samples. The other parameters are fixed from simulation. 

Four sources of backgrounds are present: the combinatorial background, the charmless background, 
the fully reconstructed (misidenfied) background and the partially reconstructed background.
The offline selection is optimized to reduce the combinatorial background contribution, and 
the remaining contamination is fitted with an exponential shape for the modes with a bachelor pion,
while it is taken flat for the \BsDK~mode. The validity of this assumption is checked with wrong-side 
samples and accounted for in the systematic uncertainty associated to the fit model. 
The other two background categories have different components in the three fits 
and therefore are explained separately. In all the fits the partially reconstructed background shapes 
are obtained fitting a non-parametric function on samples of simulated events generated in the specific exclusive modes, 
corrected for the observed mass shifts, momentum spectra, and particle identification efficiencies observed in data 
when it is needed. The yields are left free when possible or a gaussian constraint is applied if an expected 
amount is computable. 

In the \BdDp~mass fit the two relevant sources of partially reconstructed background 
are the $B^0 \to D^{*-}\pi^+$ and $B^0 \to D^{-}\rho^{+}$ decays and their
yields are left free to float in the fit. 

In the \BsDp~mass fit the misidentified \BdDp~shape is fixed from data using a 
reweighting procedure to account for misidentification momentum dependency \cite{fsfdHadr} . 
The number of expected events is computed from the yield obtained in the \BdDp~fit and the PID efficiency 
obtained from calibration sample. Its yield is therefore constrained to this expected value with a 10\% uncertainty. 
The $B^0 \to D^-_s\pi^+$ yield is calculated based on the $B^0 \to D^-_s\pi^+$ branching fraction \cite{PDG}, the measured LHCb value of \fsfdt~\cite{fsfdSemi}, 
and the value of the \BsDp~branching fraction \cite{DsKpreprint}. The shape used is the same of the signal \BsDp~with 
the mean position fixed from the \BdDp~fit. The partially reconstructed backgrounds relevant for this fit are 
the decays $B^0_s \to D^{*-}_s \pi^+$ and the $B^0_s \to D^-_s \rho^+$. Due to the large correlation between these two components, 
a gaussian constraint is used for the fraction of these two backgrounds. The fraction is assumed to be the same 
as in the $B^0$ case, while the variation is assumed to correspond to 20\% $SU(3)$ breaking. 

\begin{figure}[t]
  \centering
  \includegraphics[width=.35\textwidth]{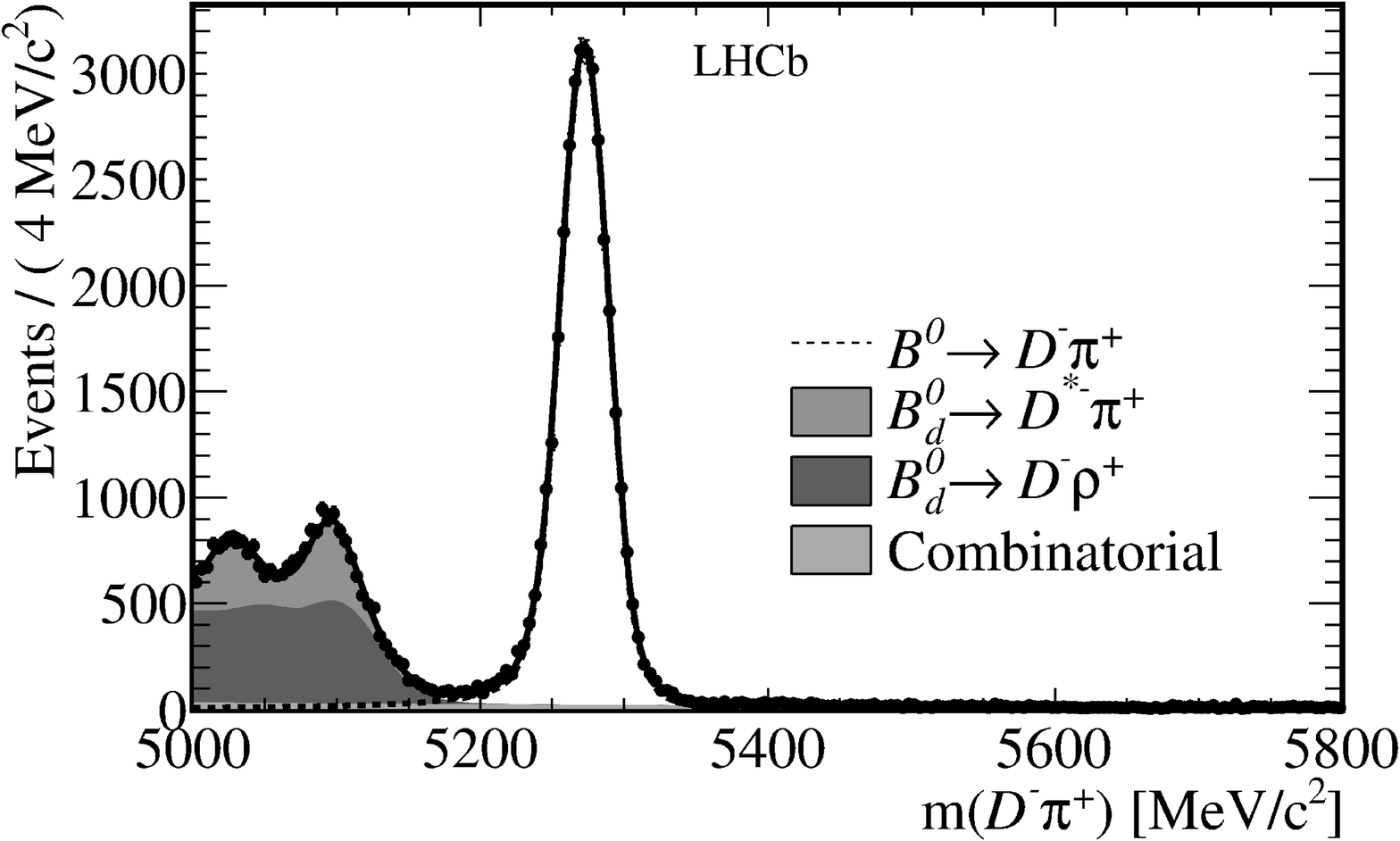}
  \includegraphics[width=.35\textwidth]{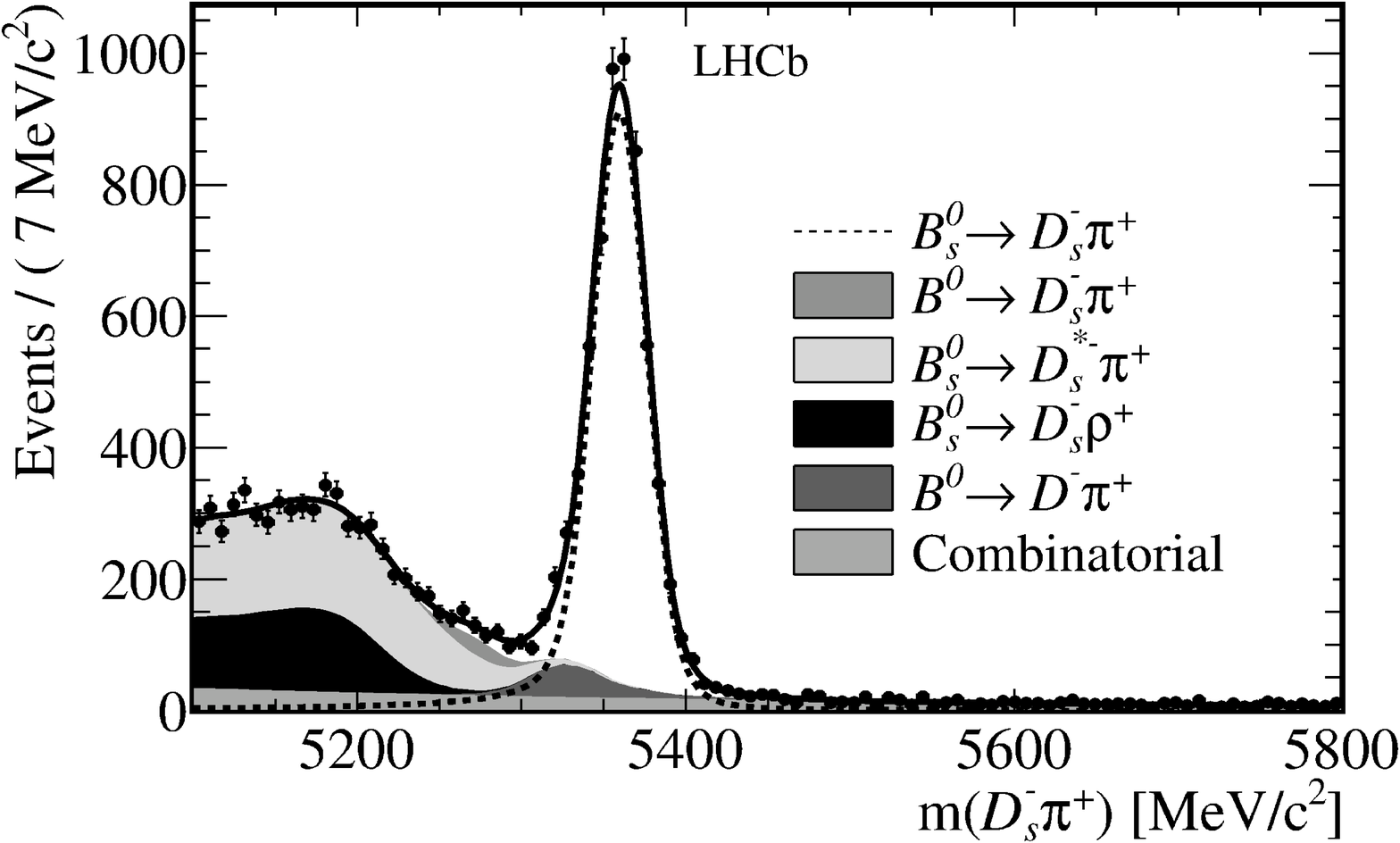}
  \includegraphics[width=.35\textwidth]{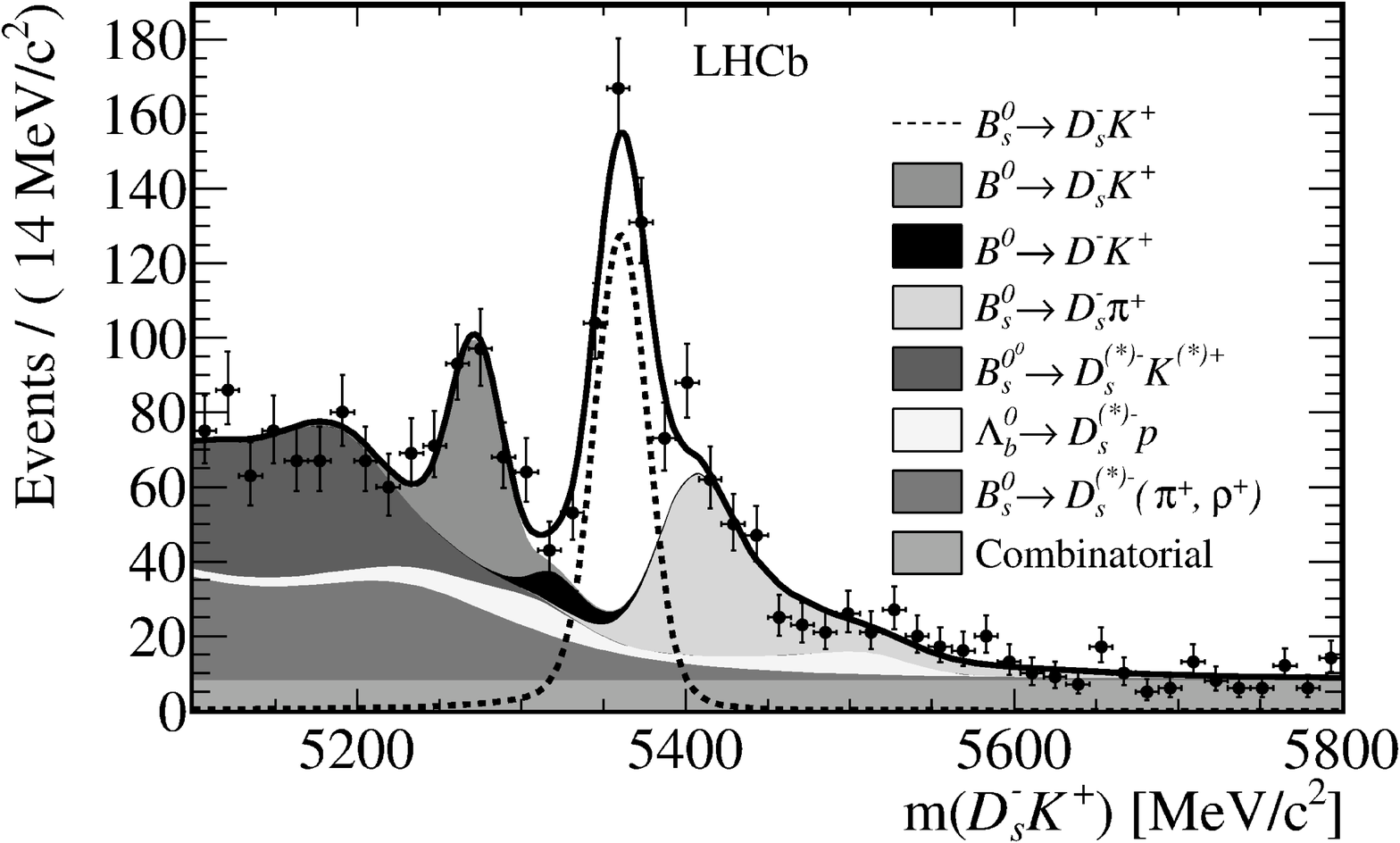}
  \caption{Mass distribution of the \BdDp~candidates (top-left), \BsDp~candidates (top-right) and \BsDK~(bottom). The stacked background shapes follow the same top-to-bottom order in the legend and in the plot. For illustration purposes the plot includes events from both magnet polarities.}
  \label{f:result}
\end{figure}

In the \BsDK~mass fit there are numerous reflections which contribute to the mass distribution. The most important reflection is 
the misidentified \BsDp~decay. Its shape is fixed from data using the reweighting procedure while the yield is left free to float. 
The same procedure is also applied on simulation sample to extract the shape of the $B^0 \to D^-K^+$ misidentified background. 
The yield is constraint according to the expected $B^0 \to D^-K^+$ yield corrected for the PID efficiency. 
In addition, there is potential cross-feed from partially reconstructed modes with a misidentified pion such as 
$B^0_s \to D_s^- \rho^+$, as well as several small contributions from partially reconstructed backgrounds with 
similar mass shapes. The yields of these modes, whose branching fractions are known or can be estimated 
are constrained to values obtained based on criteria such as relative
branching fractions and reconstruction efficiencies and PID probabilities \cite{DsKpreprint}.
The fit results are shown in Fig.~\ref{f:result}

\section{Systematic uncertainty}\label{s:systematic}
Systematic uncertainties related to the fit are evaluated by generating large sets of simulated 
experiments using the nominal fit, and then fitting them with a model where certain parameters are varied. 
The sources of systematic uncertainty considered for the fit are signal widths,
the slope of the combinatorial backgrounds, and constraints placed on specific backgrounds.
The largest deviations are due to the signal widths and the fixed slope of the combinatorial
background in the \BsDK~fit.
The systematic uncertainty related to PID 
is evaluated 
using simulated signal and calibration samples.
The observed signal yields are corrected by the difference observed 
in the (non-PID) selection efficiencies of different modes as measured from simulated
events. A systematic uncertainty is assigned on the ratio to account for percent level differences 
between the data and the simulation. These are dominated by the simulation of the hardware trigger.
A total systematic uncertainty of $3.9 \%$ for the ratio $\BsDK / \BsDp$, of $3.4 \%$ for the 
ratio $\BsDp \ \BdDp$ and of $4.6 \%$ for the ratio $\BsDK \ \BdDp$ is found. 

\section{Results}\label{s:result}
The sum of the $B^0_s \to D_s^-K^+$ and $B^0_s \to D_s^+K^-$ branching fractions relative to \BsDp~is obtained by correcting the raw
signal yields for PID and selection efficiency differences and it leads to 
\begin{equation}
\frac{BR(\BsDK)}{BR(\BsDp)} = 0.0646 \pm 0.0043 \pm 0.0025 \;,
\label{eq:Bsratio}
\end{equation}
where the first uncertainty is statistical and the second is the total systematic
uncertainty. 

The relative yields of the three decays \BdDp, \BsDp~and \BsDK~are used to extract the branching fraction of
\BsDp~and \BsDK~together with the recent \fsfdt~measurement from semileptonic decays~\cite{fsfdSemi}, leading to
\begin{eqnarray}
BR(\BsDp) &=& (2.95 \pm 0.05 \pm 0.17^{+0.18}_{-0.22})\times 10^{-3}\;,\\
BR(\BsDK) &=& (1.90 \pm 0.12 \pm 0.13^{+0.12}_{-0.14})\times 10^{-4}\;,
\end{eqnarray}
where the first uncertainty is statistical, the second is the experimental systematic 
plus the uncertainty arising from the \BdDp~branching
fraction, and the third is the uncertainty (statistical and systematic) from the
semileptonic \fsfdt~measurement. Both measurements are significantly more precise than the existing world averages~\cite{PDG}.


\section*{References}
\begin{thebibliography}{99}
\bibitem{Fleischer}R. Fleischer, \Journal{\NPB}{671}{0}{2003}.
\bibitem{Aaltonen:2008rw}T. Aaltonen {\it et al}, \Journal{PRL}{103}{19}{2009}
\bibitem{BelleBsDspi}R. Louvot {\it et al}, \Journal{\PRL}{102}{2}{2009}
\bibitem{PDG}K. Nakamura {\it et al}, \Journal{\JPG}{37}{7}{2010}
\bibitem{DetectPaper}A. A. Alves~Jr. {\it et al}, \Journal{\JINST}{3}{8}{2008}
\bibitem{DsKpreprint}R. Aaij {\it et al}, arXiv:1204.1237v1
\bibitem{TMVA}A. Hoecker {\it et al}, \Journal{\PoS}{ACAT}{040}{2007}
\bibitem{fsfdHadr}R. Aaij {\it et al}, \Journal{\PRL}{107}{21}{2011}
\bibitem{fsfdSemi}R. Aaij {\it et al}, \Journal{\PRD}{85}{3}{2012}


\end{thebibliography}
\end{document}